

\magnification=\magstep1
\baselineskip=24 true pt
\hsize=33 pc
\vsize=40 pc
\bigskip
\bigskip
\rightline {IP/BBSR-95-5 }
\rightline {January 22, 1995}
\centerline {\bf SYMMETRIES OF THE  DIMENSIONALLY REDUCED }
\centerline {\bf  STRING EFFECTIVE ACTION}
\vfil
\centerline {{\bf Jnanadeva Maharana}}
\bigskip
\centerline {\it { Institute of Physics,} }
\centerline {\it { Bhubaneswar 751 005, INDIA }}
\vfil
\centerline {\bf Abstract}

A two dimensional string  effective action is obtained by dimensionally
reducing the bosonic part of the ten dimensional heterotic string
effective action. It is shown that this effective action, with a
few restrictions on some backgrounds describes a two dimensional
model which admits an infinite sequence of nonlocal conserved currents.

\vfil
\eject


The string theory is endowed with a very rich symmetry structure.
Recently, the study of target space symmetries such as global $O(d,d)$
transformations including T-duality and S-duality, have attracted considerable
attention [1-8]. The invariance of the string effective action under
$O(d,d)$ symmetry, for the time dependent backgrounds
corresponding to string cosmological scenario, was shown in ref [2]
and subsequently the $O(d,d)$ transformation was applied to
generate new cosmological backgrounds [3,4] satisfying the equations
of motion. In a more general setting, when the backgrounds are
independent of $d$ 'internal' cordinates which are toroidally
compactified, the dimensional reduction technique can be
employed to derive the effective action as was carried out by
Schwarz and the author [6]. It was shown that the reduced action,
thus derived can be cast in a manifestly $O(d,d)$ invariant
form. The effective action  in
four spacetime dimensions has further interesting features. The field strength
of the antisymmetric tensor is dual to the axion when $D = 4$, D
being the number of spacetime dimensions. Then, the dilaton and the
axion can be combined to define a complex field. The equations
of motion associated with the four dimensional action are
invariant under an $SL(2,Z)$ transformation of this complex
field when the appropriate transformation rules are defined for
other  background fields as well. However, the effective action is not
invariant under this $SL(2,Z)$ transformation, although the equations of
motion are. Thus the four dimensional theory, obtained by the
dimensional reduction, is not only invariant under
global $O(d,d)$ transformation but also is invariant under $SL(2,Z)$
transformation. We may mention in passing that whereas T-duality is a
subgroup of the $O(d,d)$ transformation the S-duality is that of the
$SL(2,Z)$ alluded to above.

It has been observed [9-11]  recently that the three dimensional string
effective
action, dimensionally reduced from the ten dimensional heterotic string
effective action, possesses even a larger symmetry. It has been
pointed out by Sen [9] that when the ten dimensional theory is
compactified on a seven dimensional torus the resulting target
space duality group is $O(7,23;Z)$ as is expected. The target
space duality group and the strong-weak coupling duality
transformations combine into the group  $O(8,28;Z)$.
Thus we see that the four dimensional theory has
$O(d,d)$ and $SL(2,Z)$ symmetry and the three dimensional theory
has a larger symmetry group which incorporates both these
symmetries. Recently, it has been demonstrated that if the
metric, antisymmetric tensor field and the dilaton of a four
dimensional effective action depend on two of the four spacetime
coordinated then the resulting theory has an enlarged symmetry.
Bakas has shown that the dimensionally reduced theory has a symmetry
which is infinitesimally identified with the $ O(2,2)$ current
algebra [12]. The  integrable property arises due to the presence of
two commuting Killing
vectors. Furthermore there is, again, an intimate relation
between the S-duality and T-duality.

It is well known that two dimensional theories have several
novel features and possess rich symmetry structure. One of the
important attributes of a large class of two dimensional models
is the integrability property and several of these models admit
infinite sequence of nonlocal conserved currents in the
classical analysis. These conservation laws might not survive
when one considers quantum theory; however, the classical
theories have attracted considerable attention in the past.

The purpose of this note is to consider an string effective
action in two spacetime dimensions. For sake of definiteness we
consider the ten dimensional tree level effective action of the
heterotic string with graviton, antisymmetric tensor and the
dilaton as the background field. This action is dimensionally
reduced to two dimensions following the procedure adopted by
us [6]. It is expected that the two dimensional
reduced action will have interesting symmetry properties.
Indeed, the theory is endowed with an infinite number of
nonlocal conserved charges. This is our main result. The rest of
the paper is organised as follows: First we recall the essential
results of Maharana and Schwarz [6] to derive the two dimensional
effective action. Then, the coset reconstruction of the
effective action is discussed. It is essential to go over to
this formulation to demonstrate the existence of  nonlocal conserved
currents ( NCC ). Then we explicitly construct these currents
and show the conservation laws following appropriate techniques.

Let us recall how one derives the string effective action. If we
consider  the evolution of the string in the background of
its massless excitations and require that the first quantised
world sheet action respects
conformal invariance, then the
$\beta$-functions
associated with the backgrounds must vanish. In other
words,these conditions turn out to be the equations of motion to be
satisfied by the backgrounds to ensure conformal
invariance of the theory.
The tree level string effective action, involving
only the massless excitations, can be constructed in such a way that
corresponding  equations of motion of this effective action exactly
coincide with the vanishing of the $\beta$- functions.

Let us consider the bosonic part of the tree level effective
action for the heterotic string where the set of massless
background fields are graviton, antisymmetric tensor and the
dilaton. In what follows, we recapitulate the results of ref.6.
The  effective action in
 $\hat D=D+d$  dimensions ($\hat D=10$ for the present  case ) is,

$$\hat S = \int d^{\hat D}x~ \sqrt{- \hat g} e^{-\hat\phi}
\big [\hat R
(\hat g) + \hat g^{\hat \mu \hat \nu} \partial_{\hat \mu} \hat\phi
\partial_{\hat \nu} \hat\phi - {1 \over 12} ~ \hat H_{\hat \mu \hat \nu
\hat \rho} ~
\hat H^{\hat \mu \hat \nu \hat \rho}\big ].\eqno (1)$$

\noindent $\hat H$ is the field strength of antisymmetric tensor and $\hat
\phi$ is the dilaton. Here we have set all the nonabelian gauge field
backgrounds to zero. We consider the theory in a spacetime $M
\times K$, where $M$ is $D$ dimensional space and the coordinates on
$M$ are denoted by $x^{\mu}$ whereas $K$ has $d$ dimensions and
$y^{\alpha}$ are the coordinates of this submanifold. When the
backgrounds  are independent of the `internal'
coordinates $y^{\alpha}, \alpha=1,2..d$ and the internal space is
taken to be torus, the metric $\hat g_{\hat \mu \hat \nu} $
 can be decomposed as

$$\hat g_{\hat \mu \hat \nu} = \left (\matrix {g_{\mu \nu} +
A^{(1)\gamma}_{\mu} A^{(1)}_{\nu \gamma} &  A^{(1)}_{\mu \beta}\cr
A^{(1)}_{\nu \alpha} & G_{\alpha \beta}\cr}\right ),\eqno (2)$$

\noindent where $G_{\alpha \beta}$ is the internal metric and $g_{\mu\nu}$,
the $D$-dimensional space-time metric, depend on the coordinates $x^{\mu}$.
The dimensionally reduced action is,

$$ \eqalignno{ S&= \int d^Dx \sqrt {-g} e^{-\phi}
\bigg\{ R + g^{\mu \nu}
\partial_{\mu} \phi \partial_{\nu} \phi -{1\over 12}H_{\mu \nu \rho}
H^{\mu \nu \rho}& \cr
& + {1 \over 8} {\rm tr} (\partial_\mu M^{-1} \partial^\mu
M)- {1 \over 4}
{\cal F}^i_{\mu \nu} (M^{-1})_{ij} {\cal F}^{\mu \nu j} \bigg\}.
&(3) \cr}  $$

\noindent Here $\phi=\hat\phi-{1\over 2}\log\det G$ is the
shifted dilaton.

$$H_{\mu \nu \rho} = \partial_\mu B_{\nu \rho} - {1 \over 2} {\cal A}^i_\mu
\eta_{ij} {\cal F}^j_{\nu \rho} + ({\rm cyc.~ perms.}),\eqno
(4)$$

\noindent ${\cal F}^i_{\mu \nu}$ is the $2d$-component vector of field
strengths
$${\cal F}^i_{\mu \nu} = \pmatrix {F^{(1) \alpha}_{\mu \nu}
\cr F^{(2)}_{\mu \nu \alpha}\cr} = \partial_\mu {\cal A}^i_\nu - \partial_\nu
{\cal A}^i_\mu \,\, ,\eqno (5)$$
\noindent $A^{(2)}_{\mu \alpha} = \hat B_{\mu \alpha} + B_{\alpha \beta}
A^{(1) \beta}_{\mu}$ (recall $B_{\alpha \beta}=\hat B_{\alpha \beta}$), and
the $2d\times 2d$ matrices $M$ and $\eta$ are defined as
$$M = \pmatrix {G^{-1} & -G^{-1} B\cr
BG^{-1} & G - BG^{-1} B\cr},\qquad \eta =  \pmatrix {0 & 1\cr 1 & 0\cr}
\, .\eqno (6)$$

\noindent The action (3) is invariant under a global $O(d,d)$ transformation,

$$M \rightarrow \Omega^T M \Omega, \qquad \Omega \eta \Omega^T = \eta, \qquad
{\cal A}_{\mu}^i \rightarrow \Omega^i{}_j {\cal A}^j_\mu, \qquad {\rm
where} \qquad
\Omega \in O(d,d). \eqno (7)$$
\noindent and the shifted dilaton, $\phi$, remains invariant
under the $O(d,d)$ transformations. We mention in passing
that $ M \rightarrow M^{-1} $ under the duality transformation
which is a generalisation of $ R \rightarrow {1 \over R} $ duality.
Note that $M\in O(d,d)$ also and $M^T\eta M=\eta$. The
background equations of motion can be derived from the reduced
action.  The classical
solutions of the string effective action correspond to different string vacua
and are given by solutions for $M$,$\cal F$ and $\phi$. Thus,
when one obtains a set of  backgrounds satisfying the
equations of motion they correspond to a vacuum
configurations of the string theory. One can generate new
background configurations by implementing suitable $O(d,d)$
transformations on a known solution.

Let us consider the effective action in two spacetime
dimensions. We note that the term corresponding to the field
strength of the antisymmetric tensor field, $B_{\mu \nu} $, does
not contribute in $1+1$ dimensions. We assume that the dilaton
$\phi = $ constant; this assumption is made to simplify the
calculations. One can redefine the action suitably for
nonconstant dilaton as has been adopted by Bakas. Next, we set
the abelian field strength to zero from now on. Then the action
given by $ (3) $  takes the form

$$ S = \int d^2x \sqrt {-g}
\bigg\{ R +  {1 \over 8} {\rm tr} (\partial_\mu M^{-1} \partial^\mu
M) \bigg \}. \eqno (8) $$

\noindent We note that the kinetic energy term for $\phi$ does
not appear in the above equation. Since we are considering two
dimensional spacetime, we can choose the spacetime metric
$g_{\mu \nu} = e^{\alpha (x,t)} \eta_{\mu \nu} $. Here
$\eta_{\mu \nu}$ is the flat diagonal spacetime metric $=$ diag
$(-1, 1)$ ( not to be confused with the $O(d,d)$ metric ). Since
the Einstein term of the action in two dimensions is a
topological term it does not contribute to the equations of
motion. Thus the equations of motion associated with the matrix
M is of primary importance to us. The present form of the action
is not the most suitable one to derive the infinite set of NCC.
Thus we rewrite the above action in a slightly different form
following the approach of Maharana and Schwarz [6].

The matrix M introduced in equation (6) can be expressed as

$$ M_{ij} = (V^T V)_{ij} = \delta^{AB} V_{Ai}V_{Bj} \eqno (9) $$

\noindent The matrix V that produces desired M-matrix is

$$ V = \pmatrix { E^{-1} & -E^{-1}B \cr 0 & E\cr} \eqno (10) $$

\noindent where E is a $d \times d$ vielbein satisfying $E^{T}E =
G$. Moreover, V is also an $O(d,d)$ matrix since $V^{T} \eta V =
\eta$ . We would like to construct an action which is manifestly
invariant under the global $O(d,d)$ symmetry and local $O(d)
\otimes O(d)$ symmetry transformation. The desired Lagrangian
density  is

$$ {\cal L} = { 1 \over 4} \eta^{ij} \eta^{AB} (D_{\mu}V)_{Ai}
(D^{\mu}V)_{Bj}  \eqno (11) $$

\noindent Here V is an arbitrary $O(d,d)$ matrix and not of the
special form given by eq.(9). Since the $O(d,d)$ metric $\eta$
is off-diagonal form, it is not very convenient to handle the
covariant derivatives appearing in the Lagrangian above. We can
easily go over to the $O(d,d)$ metric which has $d$ diagonal
$+1$ entries $d$ diagonal $-1$ entries; denoted by $\sigma$. The
transformation matrix , $\rho$ that takes us from $\eta$ basis
to the $\sigma$  basis is given by

$$ \rho = {1 \over {\sqrt 2}} \pmatrix { 1 & -1 \cr 1 & 1\cr},
and, \sigma = \pmatrix {1 & 0\cr 0 & -1\cr } \eqno (12) $$

\noindent Now the matrix V gets rotated to $W = \rho^{T}V \rho $.
Recall that $ V^{T} \eta V = \eta $  since V belongs to $O(d,d)$,
the analogous condition satisfied by W is $W^{T} \sigma W =
\sigma $. The covariant derivative acting on W fields is defined
to be

$$ (D_{\mu}W)_{Ai} = \partial_{\mu}W_{Ai} + \omega_{\mu
AB}\sigma^{BC}W_{Ci}  \eqno (13) $$

\noindent where the $O(d) \otimes O(d)$ gauge fields
$\omega_{\mu}$ are given by

$$ \omega_{\mu} = \pmatrix { \omega^{(1)}_{\mu} & 0 \cr 0 &
\omega^{(2)}_{\mu} \cr }  \eqno (14) $$

The Lagrangian density, expressed in terms of the W fields, takes the form

$$ {\cal L} = { 1\over 4} \sigma^{ij} \sigma^{AB}
(D_{\mu}W)_{Ai}(D^{\mu}W)_{Bj}   \eqno (15) $$

\noindent The gauge fields are antisymmetric in their internal
indices. Notice that the kinetic energy term for the gauge field
does not appear in the Lagrangian and therefore, their equations
of motion are merely constraint equations. Thus the gauge fields
can be expressed in terms of $W_{Ai}$ as

$$ \omega_{\mu AB} = {1 \over 2} ( W_{Ai} \partial_{\mu} W^i_{B}
- \partial_{\mu} W_{Ai} W^i_{B} )  \eqno (16) $$

\noindent Now, substituting the expression for the gauge fields
in the Lagrangian and using the specific form of V matrix in
that Lagrangian one shows that the corresponding action is equal
to $ {1 \over 8} tr (\partial_{\mu} M^{-1} \partial^{\mu} M) $.
This proves  that the moduli $G$ and $B$ parametrise an $ O(d,d)
\over { O(d) \otimes O(d)} $ coset [6,13].

The equations of motion associated with the fields $W_{Ai}$ are

$$ (D^{\mu}D_{\mu})_{AB} W^i_{C}  \sigma^{BC} - W^k_{C}
 \sigma^{CD} W^i_{D} (D^{\mu} D_{\mu})_{AB}  \sigma^{BE}W_{Ek} = 0
\eqno (17)  $$

\noindent As is evident $i,j,k ..$ are the $O(d,d) $ index and
$A,B,C ..$ are the $O(d) \otimes O(d)$ indices. Some care must
be
exercised in deriving the equations of motion since W
is a matrix satisfying the constraint $ W^{T} \sigma W  =
\sigma $. Therefore, the equations of motion for the W fields
are not just the free field equations.

We proceed to define an $O(d,d)$ gauge potential which will play
a crucial role in deriving the set of conserved currents.
Define

$$ {{\cal A}}_{\mu ij} = {1 \over 2} ({ W_{Ai}} D_{\mu AB}
W_{Bj} - {W_{Bj}} D_{\mu AB} W_{Ai})   \eqno (18) $$

\noindent The covariant derivative appearing here is the same as
the one defined in eq.(13). We define another covariant
derivative in terms of the new gauge potential ${\cal A}_{\mu }$
as

$$ {\cal D}_{\mu} = \partial_{\mu} + {\cal A}_{\mu}   \eqno (19) $$

\noindent There are two important properties satisfied by ${\cal
A}_{\mu}$

$$ \partial^{\mu} {\cal A}_{\mu} = 0  \eqno (20) $$

\noindent is satisfied onshell; that is when the fields satisfy equations
of motion. Furthermore, the curvature associated with ${\cal A}_{\mu}$
vanishes:

$$ [ {\cal D}_{\mu} , {\cal D}_{\nu} ] = 0  \eqno (21)  $$

Now we are in a position to demonstrate the existence of the
nonlocal conserved currents [14-16]. The proof is by induction. Let us
assume that the $n^{th}$ conserved current, $ J_{\mu}^{(n)}$,
exists. Thus the current ( since we are considering two
dimensional spacetime) can be expressed as

$$ J^{(n)}_{\mu} = \epsilon_{\mu \nu} \partial^{\nu} \chi^{(n)}
  \eqno (22) $$

\noindent and the current is conserved by construction. Notice
that $J^{(n)}(x,t)$ and $\chi^{(n)}(x,t)$ carry $O(d,d)$ indices
${ij}$ and are matrix valued objects; we suppress the indices
here and everywhere. The next
level current $ J^{(n+1)}_{\mu}$ is constructed as follows:

$$ J^{(n+1)}_{\mu} = {\cal D}_{\mu} \chi^{(n)}, n \geq 0 \eqno
(23)   $$

Let us choose $ \chi^{(0)} = 0 $. Then the first current

$$ J^{(1)}_{\mu} = {\cal D}_{\mu} \chi^{(0)} = {\cal A}_{\mu}
\eqno (24)  $$

\noindent This current is conserved by the equations of motion
for ${\cal A}_{\mu}$, eq.(20).
Now we would like to show that the $ (n+1)^{th}$ current
constructed by the ansatz eq.(22) is  conserved and the proof
proceeds as follows:

$$ \partial^{\mu} J^{(n+1)}_{\mu} = \partial^{\mu} {\cal
D}_{\mu} \chi^{(n)} = {\cal D}_{\mu} \partial^{\mu} \chi^{(n)}
 \eqno (25) $$
The last equality holds due to the fact that $ [ \partial^{\mu},
{\cal D}_{\mu} ] = 0 $ as a consequence of the equation of motion
eq.(20). Note, however that, $\partial^{\mu} \chi^{(n)} =
\epsilon^{\mu \nu} J^{(n)}_{\nu} $ from eq.(22) and we can use
the relation

$$ J^{(n)}_{\nu} = {\cal D}_{\nu} \chi^{(n-1)} , n \geq 1 \eqno
(26) $$

\noindent in eq.(23) to arrive at

$$ \partial^{\mu} J^{(n+1)}_{\mu} = \epsilon^{\mu \nu} {\cal
D}_{\mu} {\cal D}_{\nu} \chi^{(n-1)}  \eqno (27)  $$

\noindent The right hand side of the equation vanishes since
 ${\cal A}_{\mu}$ is curvatureless. Thus we have
shown that the current $ J^{(n+1)}_{\mu}$ is conserved and we
can construct an infinite sequence of conserved nonlocal
currents through this procedure.

There are conserved charges $Q^{(n)}$ associated with each current
and the charge is the space integral of the time component of
the current

$$ Q^{(n)} = \int_{-\infty}^{+\infty} dx J_0^{(n)}(x,t)  \eqno (28) $$

\noindent For example,

$$ Q^{(1)} = \int_{-\infty}^{+\infty} dx {\cal A}_0 \eqno(29) $$

\noindent whereas the second charge has the form

$$ Q^{(2)} = \int_{-\infty}^{+\infty} dx {\cal D}_{0} \chi^{(1)}
\eqno (30) $$

\noindent since $J^{(2)}_{0}$ can be related to $\chi^{(1)}$
through eq.(23). Thus,

$$ Q^{(2)} = - \int_{-\infty}^{+\infty} dx J^{(1)}_{1}(x,t) +
\int_{-\infty}^{+\infty} dx J^{(1)}_{0}(x,t) \int_{-\infty}^{x}
dx' J^{(1)}_{0}(x',t)  \eqno (31) $$

\noindent The first term in (31) arises from the relation
$\partial_{0} \chi^{(1)} = - J^{(1)}_1$ and the second term
comes from the product ${\cal A}_{0} \chi^{(1)}$ and
$\chi^{(1)}$ can be written as an integral over
$J^{(1)}_{0}$ as is evident from eq.(22). We can compute all
other conserved charges in following this prescription.

Thus far we have constructed the NCC in terms of $W_{Ai}$ which
belongs to $O(d,d)$. However, the fields appearing in the string
effective action correspond to special form of $V$ given by
eq.(10). Thus the relevant $W$ takes the following form

$$ W = {1 \over 2} \pmatrix { E^{-1} + E - E^{-1}B & E - E^{-1}B
\cr E - E^{-1} + E^{-1}B & E + E^{-1} + E^{-1}B \cr } \eqno (32)
$$

\noindent Therefore, all these currents can be ultimately
expressed in terms of the moduli $G$ and $B$.

There are several comments in order at this stage. We have
demonstrated that the two dimensional reduced string theory
admits an infinite sequence of conserved nonlocal currents. In
the case of the heterotic string where one includes the
$U(1)^{16}$ Abelian gauge fields in the ten dimensional
effective action the the scalars parametrise the coset $O(8,24)
\over {O(8) \otimes O(24)}$ for two dimensional spacetime. It
obvious that the technique adopted here can be extended to
construct the currents for the moduli of the heterotic string as
well. The
antisymmetric tensor field does not contribute to the two
dimensional action. However, we made an assumption
that the dilaton takes a constant value which simplified our
calculations to some extent. If the dilaton carries
spacetime dependence, how does it affect our results ? We note
that for nonconstant dilaton the factor $e^{-\phi}$ will appear
in front of the kinetic energy term of the M-matrix. We can
redefine the M-matrix as was done by Bakas when he studied the
effective two dimensional action. However, on this occasion,
conditions given by eq.(20) and (21) cannot be satisfied
simultaneously for a single vector potential constructed from W
field. These two relations were crucial in this work to
construct the set of NCC. Therefore, the present technique will
not be useful to construct such currents. On the other hand, past
experiences with supersymmetric nonlinear $\sigma$-models show
that it might be still possible to obtain an infinite sequence
of such currents. In the case of supersymmetric $\sigma$-model
the vector that is curvatureless is not conserved and the one
which is conserved is not curvatureless; nevertheless, it is
possible to obtain a Lax pair and construct an infinite set of
nonlocal conserved charges [16]. Another important issue is whether
these classical conservation laws are valid in the quantised
theory since anomalies might creep in when we define the
currents with an appropriate normal ordering prescriptions. We
hope to report our results on some of these problems in future.

To conclude, we have demonstrated the existence of an infinite
set of nonlocal conserved current for the action under
consideration.

{\bf Acknowledgments:} I am grateful to John H. Schwarz and
Ashoke Sen for very valuable discussions at the initial stage of
this work. It is a pleasure to thank Alok Kumar for discussions
and for carefully reading the manuscript.

\vfil
\eject

\def \np {{\it Nucl. Phys. }}
\def \pl {{\it Phys. Lett. }}

\def \pr {{\it Phys. Rev. }}

\noindent {\bf References:}

\item {[1]} K. Kikkawa and M. Yamasaki, {\it Phys. Lett.} {\bf 149B}
(1984) 357; N. Sakai and I. Senda, {\it Prog. Theor. Phys.} {\bf 75}
(1986) 692; T. Busher, {\it Phys. Lett.} {\bf 194B} (1987) 59;
{\it Phys. Lett.} {\bf 201B} (1988) 466;
{\it Phys. Lett.} {\bf 159B} (1985) 127; V. Nair, A. Shapere,
A. Strominger, and F. Wilczek, {\it Nucl. Phys.} {\bf B287} (1987) 402;
A. Giveon, E. Rabinovici,
and G. Veneziano, {\it Nucl. Phys.} {\bf B322} (1989) 167;
M. J. Duff, \np {\bf B335} (1990) 610,
 A. A. Tseylin and C. Vafa, {\it Nucl. Phys.} {\bf B372}
(1992) 443;
A. A. Tseytlin, {\it Class. Quan. Gravity} {\bf 9} (1992) 979.
\item {[2]} For recent
reviews and detailed references see A. Sen Int. J. Mod. Phys.
{\bf A9} (1994) 3707;  A. Giveon, M. Porrati
and E. Ravinovici, Phys. Rep.,{\bf C244} 1994) 77.
\item {[3]} K. A. Meissner and G. Veneziano, {\it Phys. Lett.} {\bf 267B}
(1991) 33; G. Veneziano, {\it Phys. Lett.} {\bf 265B} (1991) 287,
 K. A. Meissner and G. Veneziano, {\it Mod. Phys. Lett.} {\bf A6}
(1991) 3397.
\item {[4]} M. Gasperini, J. Maharana and G. Veneziano, {\it
Phys. Lett.} {\bf 272B} (1991) 167;  M. Gasperini. J. Maharana
and G. Veneziano, {\it Phys. Lett.} {\bf 296} (1992) 51.
\item {[5]} A. Sen, \pl {\bf 271B} (1991) 295;
A. Sen, \pl {\bf 274B} (1991) 34;
S. F. Hassan and A. Sen, \np {\bf B375} (1992) 103; S. P.
Khastgir and A. Kumar, Mod. Phys. Lett. {bf A6} (1991) 3365, S.
Kar, S. P. Khastgir and A. Kumar, Mod. Phys. Lett. {\bf A7}
(1992) 1545; J. Maharana, \pl {\bf 296B} (1992) 65.
\item {[6]} J. Maharana and J. H. Schwarz, \np {\bf B390} (1993) 3. For earlier
work see S. Ferrara, C. Kounnas and M. Porrati, \pl {\bf B181} (1986) 263.
\item {[7]} J. Maharana Preprint Isaac Newton Institute for
Mathematical Sciences, Cambridge NI94023, October 1994, (Phys.
Lett.B in press).
\item {[8]} A. Font, L. Ibanez and F. Quevedo, \pl {\bf 249B}
(1990) 35; S.-J. Rey, \pr {\bf D43} (1991) 526; A. Sen \np {\bf
B404} (1993) 109; \pl {\bf 303B} (1993) 22; {\ Mod.Phys. Lett.}
{\bf A8} (1993) 2023; J. H. Schwarz, Caltech Preprint
CALT-68-1815; J. H. Schwarz and A. Sen \pl {\bf 329B}
(1994) 105, A. Sen, \pl {\bf 329B} (1994) 217
; J. H. Schwarz and A. Sen \np {\bf B411} (1994) 35.
\item {[9]} A. Sen preprint TIFR-TH-94-19.
\item {[10]} G. Segal ( to appear).
\item {[11]} E. Witten (unpublished).
\item {[12]} I. Bakas \np {\bf B428} (1994) 374.
\item {[13]} K. S. Narain \pl {bf 169B} (1986) 41; K. S. Narain,
H. Sarmadi and E. Witten \np {bf B279} (1987) 369. These works
considered toroidal compactification with spacetime independent
moduli and showed that they parametrise the coset $O(d,d) \over
{O(d) \otimes O(d)}$.
\item {[14]} M. Luscher and K. Pohlmeyer, \np {\bf B137}
(1978) 46; E. Brezin, C Itzykson, J. Zinn-Justin, J. B. Zubber,
\pl {\bf 82B} (1979) 442; H. J. de Vega, \pl {\bf 87B} (1979)
233; A. T. Ogielski, \pr {\bf D21} (1980) 3462; E. Witten, \pr
{\bf D16} (1978) 2991; P. di Vecchia and S. Ferrara, \np {\bf
B130} (1977) 93; E. Cremmer and J. Scherk, \pl {\bf 74B} (1978)
341; B. Zumino, \pl {\bf 87B} (1979) 203; E. Corrigan and C. K.
Zachos, \pl {\bf 88B} (1979) 273; T. Curtright, \pl {bf 88B}
(1979) 276; Y. Y. Goldschmidt and E. Witten \pl {\bf 91B} (1980)
392;  T. Curtright and C. K. Zachos, \pr {\bf D21} (1980)
411; H. Eichenherr and M. Forger, \np {\bf B156} (1979) 381; A.
J. Mcfarlane, \pl {\bf 82B} (1979) 239; M. Dubois-Violette and
Y. Georgelin, \pl {\bf 82B} (1979) 251.
\item {[15]} J. Maharana, \pl {\bf 128B} (1983) 411.
\item {[16]} J. Maharana, Lett. Math. Phys. {\bf8} (1984) 284;
J. Maharana, Ann. Inst. Henri Poincare {bf45} (1986) 231

\vfil
\end